\documentclass[lettersize,journal]{IEEEtran}
\usepackage{amsmath,amsfonts}
\usepackage{algorithmic}
\usepackage{algorithm}
\usepackage{array}
\usepackage[caption=false,font=normalsize,labelfont=sf,textfont=sf]{subfig}
\usepackage{textcomp}
\usepackage{stfloats,booktabs}
\usepackage{url}
\usepackage{verbatim}
\usepackage{graphicx}
\usepackage{cite}
\begin{document}

\title{{\sc AutoGuardX}: A Comprehensive Cybersecurity Framework for Connected Vehicles}

\author{Muhammad Ali Nadeem, Bishwo Prakash Pokharel, Naresh Kshetri, %,~\IEEEmembership{Senior Member,~IEEE}, 
Achyut Shankar, %~\IEEEmembership{Member,~IEEE}, 
Gokarna Sharma %,~\IEEEmembership{Senior Member,~IEEE}
\thanks{M.A. Nadeem and B.P. Pokharel are with the triOS College of Business, Technology and Healthcare, Toronto, Canada. 
\textit{\{ali.nadeem@trios, bishwo889@hotmail\}.com}}
\thanks{N. Kshetri is with the Department of Cybersecurity, Rochester Institute of Technology, Rochester, New York, USA. 
\textit{uptecnaresh@gmail.com}}
\thanks{A. Shankar is with the School of Computer Science, Engineering and Technology, Bennett University, Greater Noida, India.
\textit{ashankar2711@gmail.com}}
\thanks{G. Sharma is with the Department of Computer Science, Kent State University, Kent, Ohio, USA
\textit{gsharma2@kent.edu}}
\thanks{Corresponding author's name and email: Naresh Kshetri, \textit{uptecnaresh@gmail.com}}
%\thanks{Manuscript received 26 August, 2025; revised XXX.}
}

%\markboth{Internet of Things Journal} %

%\sloppy
\maketitle

\begin{abstract}
The rapid integration of Internet of Things (IoT) and interconnected systems in modern vehicles not only introduced a new era of convenience, automation, and connected vehicles but also elevated their exposure to sophisticated cyber threats. This is especially evident in US and Canada, where cyber-enabled auto theft has surged in recent years, revealing the limitations of existing security measures for connected vehicles. In response, this paper proposes {\sc AutoGuardX}, a comprehensive cybersecurity framework designed specifically for connected vehicles. {\sc AutoGuardX} combines key elements from existing recognized standards for vehicle security, such as ISO/SAE 21434 and ISO 26262, with advanced technologies, including machine learning-based anomaly detection, IoT security protocols, and encrypted communication channels. The framework addresses major attack vectors like relay attacks, controller area network (CAN) bus intrusions, and vulnerabilities introduced by emerging technologies such as 5G and quantum computing. {\sc AutoGuardX} is extensively evaluated through security simulations across a mix of Sedans and SUVs from four major vehicle brands manufactured between 2019 and 2023. The results demonstrate the framework’s adaptability, scalability, and practical effectiveness against existing and emerging threats. 
\end{abstract}

\begin{IEEEkeywords}
Automotive security,  Anomaly detection, Relay attacks, CAN bus intrusion, Vehicle theft
\end{IEEEkeywords}

\section{Introduction}
\IEEEPARstart{W}{ith} the rise of Internet of Things (IoT) technologies,  vehicles are becoming increasingly connected,  creating opportunities for cyber criminals to exploit vulnerabilities of such technologies. 
Auto
theft increased by 48.2\% in Ontario, Canada between 2021 and 2023 \cite{ref1}. In US, the increase was 105\% in 34 major cities between 2019 and 2023 (29\% increase in 2023 compared to 2022) \cite{ref2}. 
According to the European Union Agency for Cybersecurity (ENISA), cyber attacks on vehicles equipped with IoT technologies have increased by 62\% between 2019 and 2024 \cite{ref3}. 
The alarming rise in cyber-enabled auto theft, particularly in technologically advanced regions such as US and Canada, has exposed the limitations of conventional vehicle security systems, such as physical locks and alarms, as they are proving insufficient in preventing theft, particularly because vehicles are becoming more reliant on sophisticated electronics and digital systems. The thefts particularly exploited connectivity features such as remote keyless entry, autonomous driving systems, and in-vehicle networking. % to steal the vehicles. 

Auto theft is a major problem which not only results in direct financial loss but also affects insurance premiums, disrupts supply chains, fuels organized crime syndicates, places a significant burden on law enforcement agencies, and complicates public safety initiatives. %which often target high-demand regions such as Ontario and Quebec. Internationally, 
Studies show that stolen vehicles, particularly from Canada, 
are frequently smuggled into markets in South America, Africa, and the Middle East, making recovery impossible  \cite{ref4}. %, further complicating recovery efforts [3].  In addition to direct financial losses, the rise in auto theft drives up insurance premiums and places a significant burden on law enforcement agencies, complicating recovery efforts and public safety initiatives.

This paper introduces {\sc AutoGuardX}, a comprehensive, adaptive, and scalable cybersecurity framework designed specifically to safeguard modern connected vehicles. Unlike traditional models that address safety and security in isolation, {\sc AutoGuardX} offers a unified solution that incorporates real-time threat detection based on machine learning (ML), IoT security, and encrypted communication protocols to address these evolving threats, while also aligning with established automotive standards.  %Unlike existing solutions, {\sc AutoGuardX} offers a unified approach that adapts dynamically to emerging risks such as 5G vulnerabilities and quantum computing. 

We employ a systematic methodology consisting of four phases, Phase 1-4, in our design of the {\sc AutoGuardX} framework. Phase 1 involves an in-depth analysis of prevailing cybersecurity vulnerabilities in the automotive sector, focusing on attack techniques such as relay attacks, Controller Area Network (CAN) bus injection, and key fob cloning. This assessment was guided by case studies and industry reports, with a particular emphasis on trends in US and Canada, where sophisticated auto theft tactics are on the rise. Phase 2 reviews existing industry standards - namely ISO/SAE 21434 for cybersecurity engineering and ISO 26262 for functional safety -- to identify deficiencies in addressing modern threats at the intersection of safety and cybersecurity. 
Phase 3 evaluates the limitations of existing cybersecurity frameworks. Finally, Phase 4 proposes {\sc AutoGuardX} as a comprehensive, scalable solution for securing modern connected vehicles.

Elaborating little further,
Phase 1 systematically documents real-world theft techniques mapping them to specific vehicle vulnerabilities and attack vectors using different threat modeling approaches. Phase 2 includes a comparative compliance analysis between Original Equipment Manufacturer (OEM) security implementations and international regulatory requirements, revealing critical security gaps that current regulations do not address. Phase 3 synthesizes findings from Phases 1 and 2 to highlight the fragmented nature of current defenses, noting that existing solutions typically address either physical security or cybersecurity in isolation but not their integration. Phase 4 involves building the {\sc AutoGuardX} prototype and validating it in a controlled simulation environment replicating CAN bus networks, radio frequency (RF) keyless systems, and IoT-based vehicle communications. Testing was performed using penetration testing tools, simulated theft devices, and network intrusion methods, ensuring that the framework's capabilities were benchmarked against both industry standards and known real-world attack scenarios. 

\vspace{2mm}
\noindent{\bf Paper Organization.} The rest of paper is organized as follows.
Section \ref{section:background} provides an overview of the current landscape of auto theft, particularly in US and Canada, emphasizing the evolution from traditional theft methods to advanced cyber-enabled compromises, particularly in the context of modern connected vehicles. %Section \ref{} explores the major technical attack vectors that adversaries exploit to target connected vehicles, including relay attacks, CAN bus injection, and onboard diagnostic (OBD) port manipulation. 
Section \ref{section:prevention} reviews existing traditional and automotive cybersecurity frameworks for auto theft prevention such as ISO/SAE 21434 and Automotive SPICE and identifies the key limitations of these approaches motivating the need for a more integrated and adaptive approach. 
Section \ref{section:AutoGuardX} proposes the novel {\sc AutoGuardX} framework, outlining its architecture and core components, implementation requirements, integration with vehicle systems, and deployment phases.  %with  including real-time machine learning-based anomaly detection, secure communication protocols, IoT security mechanisms, and incident forensics capabilities. 
Section \ref{section:validation} describes the validation of %implementation and deployment methodology for 
{\sc AutoGuardX} through controlled simulation experimentation under real-world automotive cyber threat conditions. 
%covering critical aspects such as phased deployment planning, hardware and software infrastructure requirements, system integration, and testing procedures to ensure reliability and security in real-world automotive environments. 
Section \ref{section:challenge} discusses challenges and future directions to  deploy {\sc AutoGuardX} at scale.  Finally, Section \ref{section:conclusion} concludes the paper with a short discussion.

\section{Auto Theft Landscape}
%Background and Motivation}
\label{section:background}
%\subsection{Auto Theft: Landscape, Traditional, and Recent Methods}
%\subsection{Auto Theft Landscape}
Auto theft is considered a low-risk, high-reward crime. Organized criminal networks have targeted regions in Canada, such as Ontario and Quebec, as a major supply of vehicles. In US, the crime networks have targeted different cities around the country, big and small.  These criminal networks have grown increasingly confident in their ability to target vehicles. The stolen vehicles were either shipped in sea containers to South or Central America, Africa, Europe, and the Middle East or have their individual vehicle identification numbers (VINs) forged and sold within US and Canada. 

According to the Insurance Bureau of Canada, car theft was deemed a ``national crisis'' in Canada \cite{ref5}. In 2024 alone, auto insurance companies in Canada  had to pay out over \$1 billion US dollars (£860 million) in claims related to vehicle theft \cite{ref6}. Some Canadians have taken matters into their hands and employed either private neighborhood security firms or placed trackers on their vehicles. Some Canadians have even put retractable safety bollards in their drives, resembling those found at banks and embassies, to discourage burglars. Police departments across Canada have been compelled by the issue to publish public bulletins on auto theft prevention methods. The most recent vehicle theft data across the world shows that Canada has a greater theft rate (262.5 per 100,000 people) and the theft rate is 220 per 100,000 people in England and Wales \cite{ref7}. A recent study also shows that the global car shortage brought on by the COVID-19 pandemic has contributed to this increase in auto theft \cite{ref8}.

\subsection{Traditional Theft Methods}

The most traditional method is to physically break into the vehicle and attempt to start it manually \cite{ref9}. Another increasingly common method involves targeting the headlight assembly, typically the left headlight. In this approach, attackers either drill into or damage the headlight to gain access to the vehicle's wiring harness. This harness is often connected to the CAN bus -- the internal communication system linking the vehicle's Electronic Control Units (ECUs). By tapping into the CAN lines through the headlight wiring, attackers can connect a malicious device to inject fraudulent CAN messages -- such as ``unlock doors'' or ``disable immobilizer'' -- directly into the Body Control Module (BCM). This method, known as a CAN injection attack, bypasses the key fob-based authentication. Once connected, the attackers are automatically recognized as a trusted component within the vehicle network. This unauthorized access grants them control over several vehicle functions: start the engine, toggle the alarm, unlock the doors, etc. -- without the need of traditional keys or authentication. The third traditional method is to steal a key fob  and replicate it. 

The fourth traditional method is to use a device that can mimic the functions of a fob key. That device would attempt every code to locate one that unlocks the selected car model. Some thieves even resort to brute-force techniques, where they spend extended periods -- sometimes up to 30 minutes -- trying to guess the correct code to gain access. Although time consuming, this is one of widely used methods beyond the above discussion second method of physically breaking into the vehicle.

\subsection{Recent Theft Methods}
Vehicles in the recent years are utilizing IoT technologies to facilitate autonomous driving and increase driver convenience and safety, and hence called {\em connected vehicles}. Many cars can communicate with other cars to predict traffic patterns and utilize sensors and cameras to assess road conditions in real-time. Additionally, drivers may check the fuel level and lock, unlock, and start their cars using remote access. 
Connected vehicles are becoming more autonomous, making decisions on their own collecting data employing a variety of sensors, cameras, radar, and LiDAR. %In order to minimize traffic and improve routes, they can also connect with infrastructure networks. 
Vehicle-to-vehicle (V2V) connectivity is allowing vehicles to exchange data. Vehicle-to-infrastructure (V2I) connectivity is allowing vehicles to exchange data with infrastructure components such as cameras, traffic lights, and road signs.  In these connected vehicles, the common tricks criminals use to get access to the vehicle include relay attack, onboard diagnostic (OBD) port hacking, and CAN  bus injection. We discuss their working principle separately below.

\begin{figure}[!t]
\centering
\includegraphics[width=3.2in]{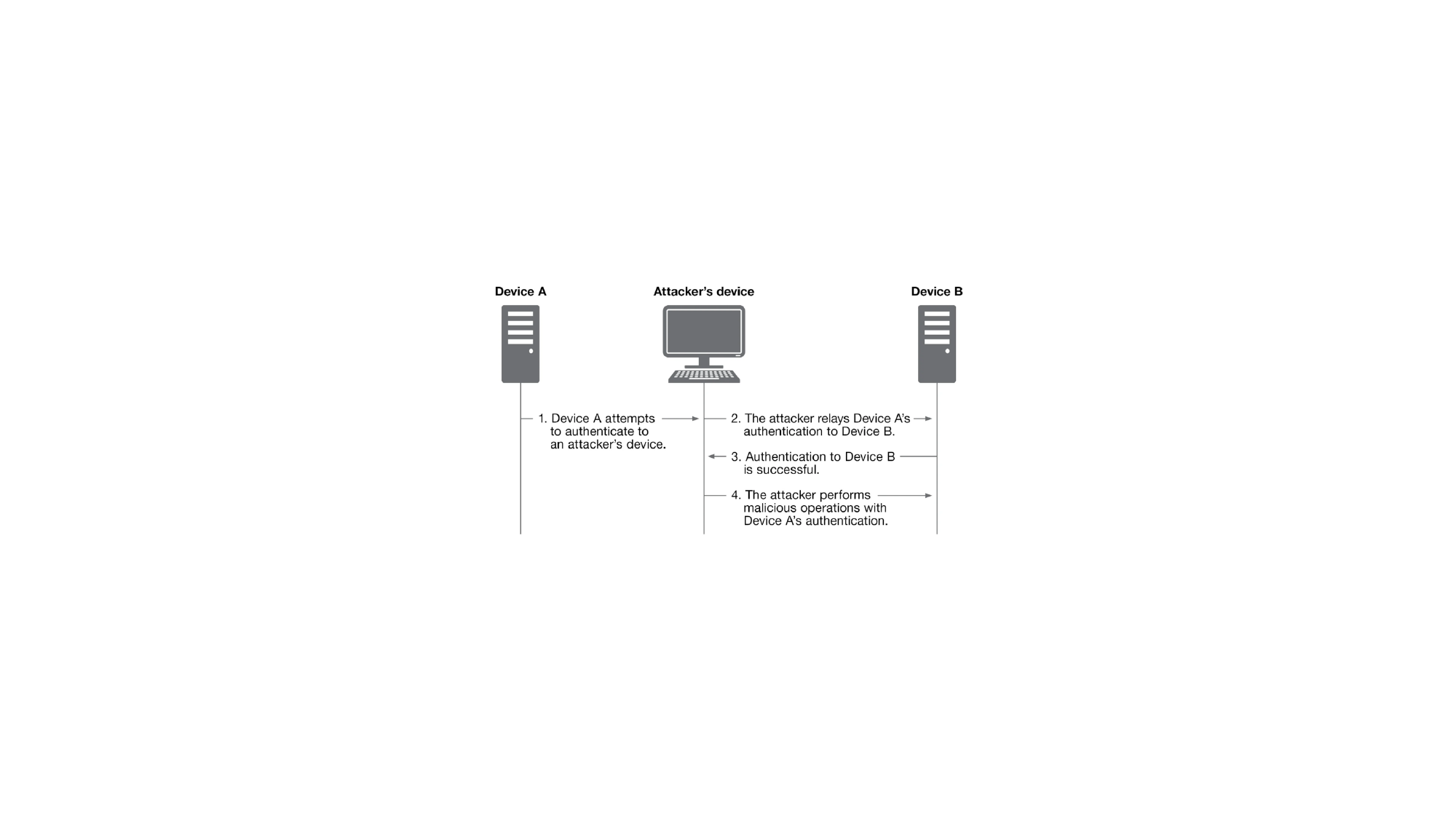}
\caption{Working principle of a relay attack.}
\label{figure:relay_attack}
\end{figure}

\begin{itemize}
\item {\bf Relay Attack:} This attack allows to take control of a car's computer by using a wireless transmitter (one open-source wireless signal hacking tool is Flipper Zero \cite{ref10}) and relay booster to intercept and relay the signal from a key fob. This methods needs the car and its key fob within a communicable distance. Thieves stand outside of homes and use transmitter devices to pick up signals coming from the key fob, amplify it to make the car think that the owner is within the communication range. 
% leaving your key close to your front door makes it easier for them to acquire the signal. Even if your wireless key fob is in your jacket or purse, the signal can still be intercepted. 
Usually, a person will be standing next to the car, ready to send the signal, open the door, and turn it on. Fig.~\ref{figure:relay_attack} illustrates the working principle of a relay attack. % {\em Experts caution that criminals are now using lock-picking equipment because they are easier to use and less expensive. Online markets readily offer these tools for less than \$60.} . 

%After gaining access to the vehicle, they typically plug a key programmer, commonly available online, into the On-Board Diagnostics (OBD) connection, usually found underneath the steering wheel, where mechanics attach a diagnostic instrument. They can then program a blank key fob to fit the car's characteristics. Studies show that this method can be used to target almost any car with a push-to-start ignition.

\item {\bf OBD Reprogramming Attack:} Another method used by criminals is to start the vehicle by plugging a device into the OBD port which is usually found underneath the steering wheel where mechanics attach a diagnostic instrument. They can then program a blank key fob to fit the car's characteristics. Studies show that this method can be used to target almost any car with a push-to-start ignition \cite{ref11}.

\item {\bf CAN Bus Injection Attack:} The CAN bus is typically located close to the front bumper. If thieves damage the connections or interfere with the signals, they can override the signals and start the car \cite{ref12}.

\item {\bf RF Signal Replay Attack:} Remote key fobs typically use RFID technology \cite{ref13}. Traditionally RFID lacked encryption making it vulnerable to replay attacks. RFID technology has evolved in recent years employing rolling codes, encryption, sophisticated modulation schemes, mutual authentication procedures, and other security-enhancing techniques \cite{ref14}. 
RF signal relay attack exploits flaw in the vehicle's RFID reader's security features, the key fob's vulnerabilities, and the communication protocol that the two employ to exchange information to gain access to the car.  
\end{itemize}

\section{Auto Theft Prevention Landscape}
\label{section:prevention}
\subsection{Traditional Theft Prevention Methods}
Traditional auto theft prevention methods primarily relied on mechanical locks and immobilizers. Mechanical locks include door locks, steering wheel locks, and gearshift locks. These devices function as purely physical barriers, requiring manual manipulation to secure or unlock the vehicle. Door locks operate through a keyed tumbler mechanism, which, while simple and cost-effective, is vulnerable to lock-picking or forced entry using basic tools. Steering wheel locks, often in the form of a metal bar fitted across the wheel, prevents the wheel from turning, thereby immobilizing the vehicle even if the ignition was bypassed. Gearshift locks prevented the gear lever from moving out of the ``Park'' or neutral position without the correct key, adding another physical hurdle for thieves. While these mechanical devices are effective deterrents against theft, they can be circumvented by skilled criminals using specialized tools or brute force \cite{ref15}. These techniques do not offer protection against modern electronic attacks.

Since 2007, Transport Canada has mandated that all passenger vehicles made or imported for sale in Canada, with a gross vehicle weight rating of 4,536 kg or less, be equipped with an engine immobilizer \cite{ref16}. An {\em immobilizer} is an electronic anti-theft device that prevents the engine from starting unless the correct, electronically coded key (or fob) is present. It works by sending a unique code from a transponder chip in the key to the vehicle's ECU. If the code does not match, the fuel system or ignition circuit remains disabled, making hot-wiring and unauthorized engine starts extremely difficult. This significantly reduced theft rates when first introduced, but modern thieves have developed advanced techniques, such as key fob cloning, relay attacks, and CAN bus injection attacks, to bypass immobilizers.

In July 2023, Transport Canada released an updated standard (ULC 338: Vehicle Theft Deterrent Equipment and System) that not only reaffirmed immobilizer requirements but also mandated aftermarket installation for older vehicles that lack them \cite{ref17}. The new standard focuses on emerging theft techniques, including CAN bus injection attacks, and recommends integrating cybersecurity features into theft deterrent systems \cite{ref17}. 

Other traditional auto theft prevention methods include car alarms which emit audible alerts when unauthorized entry is detected and vehicle identification number (VIN) etching which engraves VIN on windows to discourage resale of stolen vehicle. Physical deterrents such as wheel clamps and tire locks are also used in certain high-theft areas, but these methods, like mechanical locks, are more effective against opportunistic thefts than organized, technology-driven crimes \cite{ref18}.

\subsection{Recent Cybersecurity Frameworks}
The rise of modern connected vehicles has created new opportunities for cyber criminals to exploit vulnerabilities in vehicle systems. Traditional theft prevention methods, such as mechanical locks and immobilizers, are no longer sufficient to combat sophisticated cyber-enabled theft methods. Instead, a comprehensive framework is needed to address threats across multiple layers. %, from hardware to software to cloud-based systems.

Recent cybersecurity frameworks in the automotive sector are directly relevant to addressing the growing problem of auto theft because they establish structured approaches for mitigating both traditional and emerging cyber-physical threats. Frameworks such as ISO/SAE 21434 (Road Vehicles -- Cybersecurity Engineering) provide a lifecycle-based methodology for assessing risks and implementing safeguards across vehicle components, from embedded systems to cloud connectivity. This is particularly important in combating CAN bus injection and relay attacks, which exploit weak points in vehicle communication systems. By requiring manufacturers to integrate risk assessment, continuous monitoring, and secure software updates, ISO/SAE 21434 framework ensures that vulnerabilities are systematically identified and patched before attackers can exploit them.

Similarly, UN WP.29 R155, which governs cybersecurity management systems in vehicles, justifies the discussion by making it mandatory for automakers selling cars in regulated markets to demonstrate cybersecurity readiness. This includes countermeasures against unauthorized access via OBD ports, key fob spoofing, or wireless signal replay attacks. % - all common theft methods in Canada and the U.S. 
Additionally, the National Highway Traffic Safety Administration (NHTSA) Cybersecurity Best Practices for the Safety of Modern Vehicles (2023) emphasizes securing communication protocols (e.g., RF signals, Bluetooth, and cellular connections), logging incidents for forensic investigation, and aligning cybersecurity with functional safety standards like ISO 26262 (Road vehicles -- Functional safety). These measures not only reduce the likelihood of successful theft but also enhance recovery of stolen vehicles by generating reliable forensic data. 
In essence, these frameworks move vehicle security from a piecemeal reliance on physical deterrents to a layered cybersecurity strategy that protects against both physical tampering and digital intrusions. The standards are crucial since they provide global, enforceable standards that automotive stakeholders must adopt.
%, thereby making solutions like AutoGuardX both timely and necessary as an industry-aligned framework for combating modern auto theft.}

%{\bf GS: I think we should remove APF altogether. This para needs edit}
%The Automotive Process Framework (APF) 
We now discuss in brief several critical standards that ensure quality, safety, and security in the automotive industry. These standards encompass various domains, including software, hardware, cybersecurity, and functional safety. Table \ref{tab:comparison} shows how these standards compare to each other and to our proposed framework {\sc AutoGuardX}. 

\begin{itemize}
\item {\bf Essential SAFe 6.0 (Scaled Agile Framework):} Essential SAFe 6.0 aligns the layers of agile planning, execution, and system-level coordination for safety-critical automotive development. It promotes iterative development cycles, risk-based decision-making, and synchronized delivery across teams - enabling organizations to maintain agility while simultaneously meeting regulatory, safety, and quality standards \cite{ref19}.

\item {\bf Automotive SPICE 4.0:} Automotive SPICE (ASPICE) is the premier maturity model used to assess and improve software and hardware development processes in the automotive domain. Version 4.0, co-developed by UL Solutions’ SIS team, expands its scope by integrating cybersecurity and hardware process assessments \cite{ref19}.

\item {\bf Mechanical (Hardware) SPICE 2.0:} ME-SPICE, often branded as Mechanical SPICE, extends ASPICE methodologies to mechanical and mechatronic subsystems. It offers a framework for evaluating and enhancing development processes for mechanical components, ensuring traceable, robust engineering across electro-mechanical interfaces \cite{ref20}.

\item {\bf ISO 26262:2018 -- Road Vehicles -- Functional Safety:} ISO 26262 addresses hazards stemming from electrical/electronic failures in road vehicles. Its 2018 edition broadened its original scope (which began in 2011) to include all types of road vehicles, except mopeds \cite{ref21}.

\item {\bf ISO/SAE 21434:2021 -- Cybersecurity Engineering:} ISO/SAE 21434 defines a lifecycle-oriented cybersecurity framework for vehicles’ electrical and electronic systems - including components and interfaces. It covers the full vehicle lifecycle, from initial concept and threat analysis (TARA) to security requirement definition, through secure development, validation, production, operation, maintenance, and eventual decommissioning \cite{ref21}. Clauses concerning quality management (e.g., RQ-05-11) mandate alignment with standards such as ASPICE and other ISO quality systems. 

\item {\bf ISO 21488:2022 -- Security of the Intended Functionality (SOTIF):} ISO 21488 (formerly ISO 21448/SOTIF) addresses functional safety from a different perspective: ensuring system behavior is safe when everything is working correctly. It targets hazards arising from complex system-environment interactions like sensor misinterpretations, dynamic environmental conditions, or nominal functionality that nevertheless leads to unsafe outcomes \cite{ref22}. 
\end{itemize}

\subsection{Limitations of Existing Theft Prevention Methods}
Traditional theft prevention methods, while foundational, suffer from inherent limitations due to their static and procedural nature. Mechanical locks and alarms primarily act as deterrents but are easily bypassed with physical force or rudimentary tools. Immobilizers, although mandated in Canada since 2007, can also be compromised through techniques such as OBD port exploits or cloned keys, making them ineffective against determined attackers\cite{ref23},\cite{ref24}. Moreover, these measures lack adaptability; once criminals devise new tactics -- such as key fob cloning, CAN injection, or relay attacks --traditional systems fail to evolve to meet these threats \cite{ref25},\cite{ref26}. 

Recent cybersecurity frameworks have advanced the field, yet they too remain constrained in scope. They typically operate in silos, addressing specific life cycle phases without offering comprehensive, real-time protection. For instance, Automotive SPICE (ASPICE) provides structured development processes and requirement traceability but lacks mechanisms for runtime intrusion detection \cite{ref27}. Similarly, ISO 26262 effectively governs functional safety of electronic and electrical systems but does not address malicious cyber threats during operation \cite{ref28}. ISO/SAE 21434 focuses on cybersecurity risk management across the design and production phases, but its emphasis remains largely on process-oriented controls and post-event assessment, not continuous runtime defense \cite{ref29}. 

This segmentation exposes a critical weakness in current security practices. Modern connected vehicles are dynamic, interconnected systems that rely on multiple interfaces -- such as CAN, Ethernet, Bluetooth, and emerging 5G-based Vehicle-to-Everything (V2X) -- which can be exploited in real time. Attackers increasingly launch multi-vector intrusions, for example, combining CAN bus injection with RF relay attacks to bypass immobilizers and alarms \cite{ref30},\cite{ref31}. While current frameworks provide important governance and compliance measures, they are insufficient to detect or mitigate these attacks in the operational environments. Studies show that reliance solely on lifecycle planning or compliance audits leaves vehicles exposed to zero-day exploits and adaptive adversarial strategies \cite{ref32},\cite{ref33}.

Consequently, there is a clear need for a unified, adaptive, and runtime-focused framework that bridges these gaps. The proposed {\sc AutoguardX} framework aims to address these shortcomings by integrating real-time monitoring, anomaly detection powered by ML \cite{ref34}, IoT security measures \cite{ref35},\cite{ref36}, and encrypted communication protocols \cite{ref37}. Unlike existing approaches, {\sc AutoguardX} emphasizes continuous operational protection, forensic readiness, and adaptability to evolving threats, positioning it as a comprehensive defense solution against both current and future forms of auto theft.

\begin{table*}[!t]
\caption{Comparing {\sc AutoGuardX} and existing frameworks. SOTIF -- Safety Of The Intended Functionality \label{tab:comparison}}
\centering
\begin{tabular}{|p{.9in}|p{.8in}|p{.6in}|p{.7in}|p{.6in}|p{.7in}|p{.8in}|p{.7in}|}
\toprule
{\bf Feature} & {\sc AutoGuardX} & {\bf Essential SAFE 6.0} & {\bf Automotive SPICE 4.0} & {\bf Mechanical SPICE 2.0} & {\bf ISO 26262:2018} & {\bf ISO 21434: 2021}  & {\bf ISO 21488:2022} \\
\toprule
{\bf Real-time Detection} & 
ML + Blocking
& Iterative only
& Process only
& Mechanical only
& No runtime defense
& Lifecycle only
& Hazards only\\
\hline
{\bf IoT Device Security} 
& OTA, Authentication, Isolation
& Agile focus
& Process only
& Not covered
& Device Safety only
& ECU + IoT focus
& Environmental hazards only\\
\hline
{\bf CAN Bus Protection} &
Encrypted + Anomaly Isolation
& Risk guidance only
& Analysis advised
& Not addressed
& Error handling only
& Suggested protection
& Not specified\\
\hline
{\bf ML Integration} &
Adaptive ML Prediction
& No ML
& Not included
& Not included
& Verification only
& Not covered
& Out of scope\\
\hline
{\bf Secure Communication Protocols} & 
Encrypted RF, Rolling CAN
& Generic suggestion
& Not included
& Not included
& Safe communications only
& Secure communication required
& Not addressed\\
\hline
{\bf Incident Logging and Forensics} & 
Built-in Forensics + Reporting
& Documented process
& Process maturity
& Mechanical trace only
& Traceability only
& Logging emphasized
& Mitigation included \\
\hline
{\bf Regulatory Alignment} & 
ISO 26262 + 21434
& Safety/Cyber aligned
& Process compliant
& Mechanical compliant
& Full FS compliant
& Full cybersecurity
& SOTIF compliant \\
\hline
{\bf Platform Scalability} & 
Modular + Cross-platform
& Agile scalable
& Tailoring needed
& Tailoring needed
& ASIL-based tailoring
& System-tailorable
& Variant-applicable\\
\bottomrule
\end{tabular}
\end{table*}

\begin{figure*}[!t]
\centering
\includegraphics[width=5.5in]{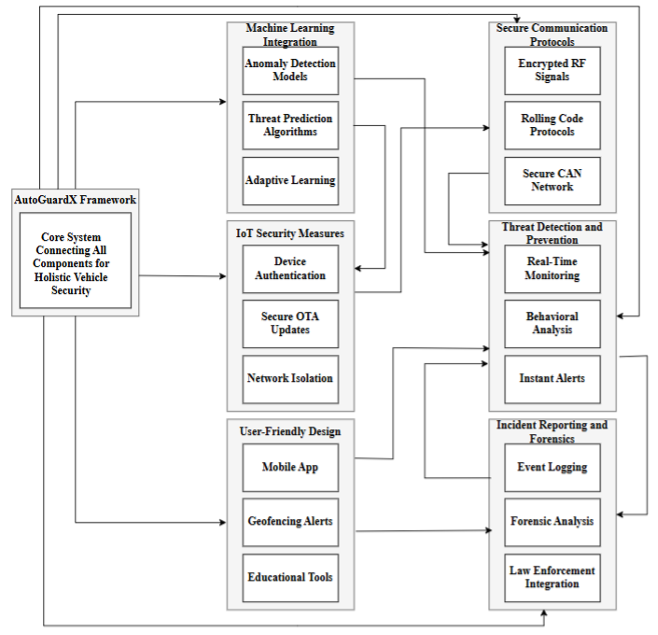}
\caption{{\sc AutoGuardX} framework.}
\label{figure:framework}
\end{figure*}

\section{AutoGuardX Framework}
\label{section:AutoGuardX}
The {\sc AutoGuardX} framework is designed to address emerging cybersecurity threats in the automotive industry, offering a holistic approach to vehicle security. 
The {\sc AutoGuardX} framework is purpose-built to address the deficiencies of the existing automotive standards, delivering an integrated solution that complements, rather than replaces, the  existing automotive standards. {\sc AutoGuardX} unites advanced technologies - machine learning, IoT security, secure communications, and forensic intelligence -- with normative structures like ISO 26262 and ISO/SAE 21434. 
Fig.~\ref{figure:framework} illustrates an architecture of {\sc AutoGuardX}. Table \ref{tab:comparison} compares {\sc AutoGuardX} with the existing frameworks. 

\begin{itemize}
\item {\bf Real-Time Threat Detection and Prevention:} Powered by ML-based anomaly detection, {\sc AutoGuardX} continuously monitors in-vehicle and network behaviors, enabling real-time detection and active blocking of threats based on historic and real-time data. These capabilities are absent in current ISO or SPICE frameworks. 
\item {\bf Secure Communication Protocols:} {\sc AutoGuardX} enforces encrypted RF links, encrypted CAN bus, and rolling-code protection for keyless entry -- areas usually left vague or unaddressed in the existing standards. 
%\item Machine Learning Integration: Adaptive threat prediction models evolve based on historic and real-time data. Unlike ISO or SPICE, which don’t integrate ML at runtime, AutoGuardX dynamically refines its defensive posture.
\item {\bf IoT Device Security:} Vehicles increasingly embed IoT devices. {\sc AutoGuardX} highlights the zero-trust principle missing from legacy standards by securing firmware with over-the-air (OTA) updates, ensuring device authentication, and isolating network components. 
\item {\bf Incident Logging and Forensics:} Departing from the existing standards, comprehensive logging and forensic tools are embedded in {\sc AutoGuardX} providing detailed event capture and analysis without relying on external processes.
\item {\bf Regulatory Compatibility and Platform Scalability:} The {\sc AutoGuardX} architecture is modular and platform-agnostic. It aligns explicitly with ISO 26262 and ISO/SAE 21434 requirements, easing certification efforts, while its plug-and-play design supports flexible deployment across different Original Equipment Manufacturer (OEM) platforms and vehicle variants.
\end{itemize}

Table \ref{tab:mitigation} illustrates how the innovations in above discussed apsects help {\sc AutoGuardX} to mitigate common cyber attacks on modern connected vehicles. 

We now discuss {\sc AutoGuardX} in details. We first discuss components of {\sc AutoGuardX}; Fig.~\ref{figure:framework} illustrates the {\sc AutoGuardX} components as well as the aspects we consider within each component. We then discuss implementation requirements and integration with vehicle systems. Finally, we discuss how to deploy {\sc AutoGuardX}.

\subsection{{\sc AutoGuardX} Components}
\subsubsection{ML Integration}
The ML integration component is central to {\sc AutoGuardX}'s capability to predict and detect cyber threats proactively. With its ability to learn from patterns and continuously adapt to new information, ML integration enhances the security and adaptability of vehicle protection. We integrate ML models tailored to various aspects.
\begin{itemize}
\item {\bf Anomaly Detection Models:} These models monitor vehicle operations to identify any deviations from normal behavior, flagging potential security risks. Studies show that anomaly detection is effective in identifying new types of threats, which is particularly crucial as cybercriminals continue to evolve their tactics \cite{ref38}. 
\item {\bf Threat Prediction Algorithms:} By analyzing historical data and patterns, these algorithms predict possible future threats, allowing the system to act before an attack occurs. This approach has been successfully implemented in various domains, including cybersecurity \cite{ref39}. 
\item {\bf Adaptive Learning Models:} The system continuously improves its threat detection models by learning from past incidents and user behavior. This adaptive learning approach allows {\sc AutoGuardX} to stay ahead of emerging threats, making it more effective in providing real-time security \cite{ref40}.
\end{itemize}

\subsubsection{IoT Security Measures}
 The IoT security measures component protects vehicle networks and ensures that devices and communication channels are secure from unauthorized access. This component incorporates several aspects.
 \begin{itemize}
\item {\bf Device Authentication:} Secure device authentication ensures that only trusted devices can connect to the vehicle’s network, preventing unauthorized access. Research in IoT security highlights the importance of device authentication and protecting systems from cyber attacks \cite{ref41}. 
\item {\bf Secure OTA Updates:} OTA updates allow the vehicle’s software to be updated remotely. These updates must be encrypted to prevent tampering. This process is vital for maintaining up-to-date security features and patching vulnerabilities in real-time \cite{ref42}. 
\item {\bf Network Isolation:} By segmenting critical vehicle systems from non-critical systems, {\sc AutoGuardX} ensures that even when an non-essential IoT device is compromised, the core vehicle system remains secure. Network isolation is a best practice in IoT security, reducing the risk of widespread attacks.
\end{itemize}

\subsubsection{Secure Communication Protocols}
Communication within and between the vehicle’s systems and external devices must be secure to protect against unauthorized access. The secure communication protocols component of {\sc AutoGuardX} provides encryption and protection for all data transmission. Particularly, this component incorporates the following aspects.
\begin{itemize}
\item {\bf Encrypted RF Signals:} RF signals, commonly used for keyless entry and remote operations, are encrypted to prevent attackers from intercepting and replaying these signals. RF signal encryption is widely regarded as a critical measure to prevent vehicle theft via relay attacks \cite{ref43}. 
\item {\bf Rolling Code Protocols:} Rolling code protocols change authentication codes with each use, ensuring that intercepted signals cannot be reused. This system is a standard method of preventing replay attacks in modern vehicle security systems \cite{ref44}. 
\item {\bf Secure CAN Network:} The CAN is integral to vehicle communications. {\sc AutoGuardX} ensures that the CAN network is secure, preventing cyber criminals from sending malicious commands to vehicle systems. Securing the CAN network is a crucial step in protecting the vehicle’s internal systems \cite{ref45}. 
\end{itemize}

\subsubsection{Incident Reporting and Forensics}
This component of {\sc AutoGuardX} ensures that any security events are logged and thoroughly analyzed. This enables quick response and improves future defense mechanisms.
Particularly, this component incorporates the following aspects.
\begin{itemize}
\item {\bf Event Logging:} {\sc AutoGuardX} continuously logs important security events, such as unauthorized access attempts or system malfunctions. Logging provides a comprehensive record that is vital for investigation and analysis.
\item {\bf Forensic Analysis:} After a security incident, forensic analysis tools examine the event data to understand how the attack occurred and what vulnerabilities were exploited. Forensic analysis helps improve future security protocols and prevent similar attacks \cite{ref46}. 
\item {\bf Law Enforcement Integration:} In the event of a theft or a cyber attack, {\sc AutoGuardX} can automatically share relevant data with law enforcement, aiding in faster response times and the recovery of stolen vehicles. Law enforcement agencies increasingly rely on real-time data sharing to combat cyber crimes \cite{ref47}. 
\end{itemize}

\subsubsection{Threat Detection and Prevention}
This component is at the core of {\sc AutoGuardX}, continuously monitoring for and mitigating security threats in real time. This component incorporates the following aspects.
\begin{itemize}
\item {\bf Real-time Monitoring:} The system actively monitors all vehicle systems and external communication channels, ensuring any signs of a cyberattack or unauthorized access are detected immediately. Real-time monitoring is a key part of proactive security in the automotive industry.
\item {\bf Behavioral Analysis:} By analyzing the normal behavior of the vehicle’s systems, {\sc AutoGuardX} can identify any anomalies that could signal a potential security threat. This behavior-based approach has proven to be effective in detecting new and previously unseen attack methods \cite{ref48}.
\item {\bf Instant Alerts:} When a threat is detected, the system sends immediate alerts to the vehicle owner or connected security services. These alerts allow for rapid response, minimizing potential damage from a cyber attack or theft attempt.
\end{itemize}

\subsubsection{User-Friendly Design}
{\sc AutoGuardX} is only effective if vehicle owners can easily interact with it. This ensures that the security features are accessible and understandable to everyday users. This component tailors to the following aspects.
\begin{itemize}
\item {\bf Mobile App:} The AutoGuardX mobile application serves as a central interface for vehicle owners, allowing them to monitor their vehicle’s security status, receive alerts, and control key features remotely. Mobile apps are increasingly being used for managing IoT-connected devices, offering both convenience and security \cite{ref49}. 
\item {\bf Geofencing Alerts:} Using GPS technology, geofencing creates a virtual boundary around the vehicle. If the vehicle moves outside this boundary, the system sends an immediate alert. This feature enhances real-time security and helps protect the vehicle from theft \cite{ref50}. 
\item {\bf Educational Tools:} {\sc AutoGuardX} includes educational materials to inform users about common security threats and how to use the system effectively. This helps ensure that users can maximize the protection provided by the system.
\end{itemize}

\begin{table*}[!t]
\caption{Common Vehicle Cyber Attacks and {\sc AutoGuardX} Mitigation Strategies\label{tab:mitigation}}
\centering
\begin{tabular}{|p{1.2in}|p{1.7in}|p{1.8in}|p{1.7in}|}
\toprule

{\bf Attack Vector} &
{\bf Attack Method} &
{\bf AutoGuardX Mitigation} &
{\bf Result} \\
\toprule
{\bf Relay} 
& RF signal relay to unlock/start vehicle
& Encrypted RF, rolling codes, geofencing alerts
& Prevents replay, triggers alerts\\
\hline
{\bf CAN Bus Injection}
& Fake message via headlight/CAN port
& CAN message authentication and encryption
& Blocks unauthorized signal injection\\
\hline
{\bf OBD Reprogramming}
& Use of key programmer on diagnostic port
& Device authentication, secure OBD access
& Prevents unauthorized fob cloning \\
\hline
{\bf RF Signal Replay} 
& Record and reuse signal from key fob
& Rolling codes and signal variance, anomaly detection
& Mitigates static replay attempts\\
\hline
{\bf Wireless Hacking Tools}
& Use of Flipper Zero/low-cost RF tools
& RF signature anomaly detection, secure communication
& Early detection and device lockdown\\
\bottomrule
\end{tabular}
\end{table*}

\begin{table*}[!t]
\caption{Hardware and Software Requirements for AutoGuardX Implementation\label{tab:hwswrequirement}}
\centering
\begin{tabular}{|p{1.5in}|p{2.2in}|p{1.0in}|p{1.8in}|}
\toprule
{\bf Component} & 
{\bf Description} & 
{\bf Requirement Level} &
{\bf Supported Technologies} \\
\toprule
Onboard Processor Unit & 
Runs ML and encryption modules & 
Critical & 
ARM Cortex, Intel Atom\\
\hline
Secure Communication Modules & 
Encrypts CAN/RF/Bluetooth traffic & 
Critical & 
AES-CTR, TLS 1.3, ChaCha20\\
\hline
IoT Gateway & 
Connects vehicle sensors to backend securely & 
Critical & 
MQTT, HTTPS, ZeroMQ\\
\hline
ML Frameworks & 
Supports anomaly detection, behavioral analysis & 
Essential & 
TensorFlow, PyTorch, Scikit-learn \\
\hline
OTA Update System & 
Secure firmware/software updates over network & 
Essential & 
Encrypted HTTPS, Code signing \\
\hline
Logging \& Forensics Tool & 
Stores and analyzes security events post-incident & 
Recommended & 
ELK Stack, Graylog, custom JSON logs\\
\bottomrule
\end{tabular}
\end{table*}

\subsection{{\sc AutoGuardX} Implemention Requirements }

The implementation of {\sc AutoGuardX} requires a robust hardware and software infrastructure to handle the vast amounts of data generated by vehicle systems and ensure that security features are delivered in real time. Table \ref{tab:hwswrequirement} illustrates such requirements for {\sc AutoGuardX} implementation, in addition to the requirement level (recommended, essential, critical) and supported technologies. 

\subsubsection{Hardware requirements} The following are the hardware requirements for implementing {\sc AutoGuardX} in vehicle systems.  
\begin{itemize}
\item {\bf Onboard Vehicle Units:} The {\sc AutoGuardX} framework relies on secure onboard hardware modules capable of running the framework’s software in real time. These modules include micro-controllers and processors for threat detection and communication encryption.
\item {\bf Communication Modules:} These modules manage the secure transmission of data between vehicle systems, external devices, and the cloud. They must support various protocols such as CAN, RF, and Bluetooth, each secured according to {\sc AutoGuardX} standards.
\item {\bf Sensors and IoT Devices:} Sensors in modern vehicles are key to detecting anomalies, monitoring the environment, and collecting data for ML models. These IoT devices require robust security measures to prevent them from being compromised.
\end{itemize}

\subsubsection{Software Requirements}
The following are the software requirements for implementing {\sc AutoGuardX} in vehicle systems.  
\begin{itemize}
\item {\bf {\sc AutoGuardX} Software Platform:} The platform integrates ML algorithms, secure communication protocols, and vehicle management systems. It should be compatible with the underlying operating system of the vehicle, including infotainment and autonomous driving systems.
\item {\bf ML Libraries:} The integration of anomaly detection models and predictive algorithms requires the use of advanced ML libraries such as TensorFlow \cite{ref51}, Scikit-learn \cite{ref52}, and PyTorch \cite{ref53}.
\item {\bf Security Protocols and Cryptographic Libraries:} For secure communication, libraries that support AES encryption, RSA key exchange, and public-key infrastructure (PKI) are essential for ensuring secure communication within the vehicle and with external devices \cite{ref54}.
\end{itemize}

\subsection{{\sc AutoGuardX} Integration with Vehicle Systems}

%C. INTEGRATION WITH VEHICLE SYSTEMS
The {\sc AutoGuardX} framework must seamlessly integrate with existing vehicle systems, including in-vehicle networks (e.g., CAN bus), infotainment, and autonomous vehicle systems. This integration is a critical part of the deployment process.

\subsubsection{In-Vehicle Network Integration} We consider following aspects for in-vehicle network integration.
\begin{itemize}
\item {\bf  CAN:} It is a foundational communication protocol in vehicles that enables various subsystems to interact. {\sc AutoGuardX} ensures that CAN messages are encrypted and verified for authenticity. Secure CAN protocols prevent unauthorized access to critical vehicle systems, reducing the risk of remote hacking.
\item {\bf Autonomous Driving Systems:} For vehicles with autonomous driving capabilities, {\sc AutoGuardX} integrates with these systems to monitor for any signs of cyber attack or malfunction. Real-time monitoring and threat prediction algorithms help ensure that autonomous vehicles remain secure against cyber threats.
\end{itemize}

\subsubsection{Infotainment and Telematics Systems}
Infotainment and telematics systems provide entry points for potential cyber attacks. {\sc AutoGuardX} ensures that these systems are isolated from critical vehicle functions to limit the impact of any vulnerabilities. Additionally, secure RF signals are used to prevent hacking attempts on wireless communication systems, such as keyless entry and remote vehicle control.

\begin{figure}[!t]
\centering
\includegraphics[width=2.8in]{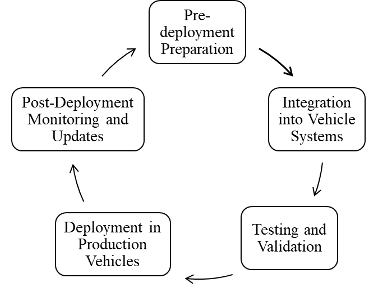}
\caption{Deployment phases of {\sc AutoGuardX} framework.}
\label{figure:deploymentphases}
\end{figure}

\subsection{{\sc AutoGuardX} Deployment}
The deployment of the AutoGuardX framework involves multiple stages to ensure that the system is seamlessly integrated into vehicle systems and operates as intended. 
We adapt the phased deployment which ensures a systematic and controlled introduction of security features into the automotive ecosystem, minimizing risks during the transition \cite{ref55}. Fig.~\ref{figure:deploymentphases} depicts the five phases of {\sc AutoGuardX} deployment.

\begin{itemize}
\item {\bf Pre-deployment Preparation:} In this phase, the vehicle manufacturer or cybersecurity team conducts a detailed assessment of the vehicle’s existing security architecture. This includes identifying key vulnerable areas, defining security objectives, and selecting appropriate hardware and software components for AutoGuardX integration.
\item {\bf Integration into Vehicle Systems:} During this phase, the AutoGuardX components are integrated into the vehicle’s control systems, communication networks (such as CAN), and IoT devices. Secure connections are established, and the framework is configured to monitor various vehicle functions, including infotainment, navigation, and autonomous driving systems.
\item {\bf Testing and Validation:} Extensive testing is conducted to ensure the system operates correctly under different conditions. Testing includes security simulations, stress testing under attack scenarios, and compatibility checks with existing systems. Additionally, the system undergoes functional testing to ensure that threat detection, communication protocols, and user interface are effective.
\item {\bf Deployment in Production Vehicles:} After successful testing, the AutoGuardX framework is deployed in production vehicles. This stage involves configuring over-the-air (OTA) updates for continuous improvement and patch management.
\item {\bf Post-Deployment Monitoring and Updates:} Continuous monitoring ensures the system remains up to date with evolving threats. OTA updates are used to enhance functionality, improve machine learning models, and address newly discovered vulnerabilities. The deployment team also collects feedback from users to address any issues promptly.
\end{itemize}

\section{{\sc AutoGuardX} Validation}
\label{section:validation}
We evaluated effectiveness and robustness of {\sc AutoGuardX}  through a series of structured tests conducted in a controlled real-time simulated environment.  We use commercially available automotive cybersecurity tools and mobile applications.  These evaluations were designed to mimic real-world threat scenarios, ensuring that the framework adheres to recognized industry standards including ISO/SAE 21434 for automotive cybersecurity engineering, UN WP.29 R155 for cybersecurity management systems, and the NHTSA Cybersecurity Best Practices for the Safety of Modern Vehicles (2023) for U.S. regulatory alignment.

%\noindent{\bf Environmental Setup.} We developed a real-time simulated threat environment. 

\subsection{Attack Simulation Testbed and Tools}
To evaluate the effectiveness of {\sc AutoGuardX} against real-world threats, we established  a testbed to capture a comprehensive simulated automotive threat environment. This environment combined attack simulation tools, hardware-in-the-loop systems, and a multi-layer software stack to replicate the conditions of modern connected vehicles. The design allowed us to test {\sc AutoGuardX} against both traditional theft techniques (e.g., OBD port exploitation, key fob spoofing) and advanced cyber intrusions (e.g., CAN injection, 5G-based DDoS). The hardware, software, and network testbed configurations are shown in Tables \ref{tab:hwtestbed}, \ref{tab:swtestbed}, and \ref{tab:nwconfiguration}, respectively.

The attack simulation toolkit incorporated a blend of commercial, open-source, low-cost, and custom solutions, ensuring realism and diversity in threat modeling. Vector CANoe (v17.0) and CANayzer (SP3) were used as the core platforms for simulating and analyzing in-vehicle communications across CAN, Ethernet, and FlexRay protocols, widely adopted in OEM architectures \cite{ref56}. These tools enabled structured injection of malicious traffic into ECUs and provided detailed timing analysis for detection validation. For testing OTA vulnerabilities, Burp Suite (2024.8) \cite{ref57} was employed to simulate man-in-the-middle (MITM) attacks on software update channels, while the Argus Cybersecurity Suite was used for benchmarking intrusion detection in automotive environments \cite{ref58}.

In addition, we use open-source tools to simulate certain attacks. We use Kali Linux (2024.3), combined with the HackRF One software defined radio (SDR), for RF interception and replay attacks, replicating how criminals exploit vulnerabilities in remote keyless entry systems operating at 315 MHz and 433 MHz. Previous studies have demonstrated the feasibility of such replay attacks using low-cost software defined ratio (SDR) devices \cite{ref59}. To replicate realistic street-level threats, we incorporated eBay-sourced tools such as the XTOOL X100 PAD key programmer, often exploited in illicit markets for OBD based vehicle reprogramming, and RF amplifiers to extend the attack range of key fob cloning attempts \cite{ref60}. Finally, custom-developed Python scripts, executed on Raspberry Pi 4 devices, were used to emulate CAN injection attacks and 5G DDoS attacks, ensuring {\sc AutoGuardX}'s resilience against emerging next-generation threats. 

The hardware testbed consisted of ECUs sourced from major OEMs, configured to support multiple communication protocols (CAN, Ethernet, and FlexRay). These ECUs were connected with RF transceivers, BLE 5.0 modules, and a 5G emulator (Keysight E7515B UXM Wireless Test Platform) to simulate vehicle-to-vehicle (V2V) and vehicle-to-infrastructure (V2I) communication threats \cite{ref61}. A range of sensors and IoT-enabled subsystems, including LiDAR (Velodyne Puck), radar, cameras, and Android Automotive OS-based infotainment systems \cite{ref62}, expanded the attack surface to mimic a fully connected vehicle ecosystem. For onboard processing, NXP S32K micro-controllers were used for CAN handling, while NVIDIA Jetson Nano processors executed {\sc AutoGuardX}'s ML models, enabling real-time anomaly detection, secure communication, and forensic event logging \cite{ref63}. 

The experimental platform is further strengthed through 
the software stack. % further strengthened this environment. 
{\sc AutoGuardX} was deployed on QNX 7.1 real-time operating system, integrating TensorFlow 2.15 and Scikit-learn 1.4 for ML-based anomaly detection, alongside AES-256 encryption and RSA-2048 key exchange for secure communications \cite{ref64}. Attack scenarios were emulated through Vector CANoe/CANalyzer simulations and Python-based adversarial scripts, while Wireshark (v4.2) continuously monitored encrypted RF, Bluetooth, and CAN traffic to validate the integrity of communications and detect anomalies \cite{ref65}. 

%By combining industrial-grade automotive cybersecurity platforms with criminal-grade hacking tools and custom adversarial scripts, the testbed offered a practical and future-oriented assessment of AutoGuardX's ability to mitigate both current theft vectors and emerging cyber-physical threats.
%}

\begin{table}[!t]
\caption{Hardware Testbed Configuration\label{tab:hwtestbed}}
\centering
\begin{tabular}{|p{.7in}|p{1.2in}|p{1.1in}|}
\toprule
{\bf Component} & 
{\bf Description} & 
{\bf Purpose}  \\
\toprule
Electronic Control Units (ECUs) & 
ECUs from four OEMs (Brands A-D) with CAN, Ethernet, FlexRay support & 
Simulate in-vehicle networks for integration and attack response testing \\
\hline
Communication Modules & 
RF (315/433 MHz), Bluetooth LE (5.0), 5G (Keysight) E7515B UXM Test Platform) & 
Replicate V2I and V2V communications for connectivity threat testing \\
\hline
Sensors and IoT Devices & 
LiDAR (Velodyne Puck), radar, cameras, telematics, infotainment (Android Auto) & 
Mimic connected vehicle setups for sensor-based threat detection \\
\hline
Onboard Processing Units & 
NXP S32K microcontrollers, NVIDIA Jetson Nano for ML tasks & 
Support real-time anomaly detection, encryption, and forensic logging \\
\bottomrule
\end{tabular}
\end{table}

\begin{table}[!t]
\caption{Software Testbed Configuration\label{tab:swtestbed}}
\centering
\begin{tabular}{|p{.7in}|p{1.2in}|p{1.1in}|}
\toprule
{\bf Component} & 
{\bf Description} & 
{\bf Purpose}  \\
\toprule
{\sc AutoGuardX} platform & 
QNX 7.1 RTOS with TensorFlow 2.15, Scikit-learn 1.4, AES-256, RSA-2048 & 
Run secure protocols, ML-based anomaly detection, and forensic logging\\
\hline
Simulation Software & 
Vector CANoe (v17.0), CANalyzer SP3, Python 3.11 scripts for attacks & 
Simulate CAN bus interactions and advanced attack vectors \\
\hline
Network Monitoring & 
Wireshark (v4.2) for RF, Bluetooth, CAN traffic analysis & 
Validate encryption integrity and detect communication anomalies \\
\bottomrule
\end{tabular}
\end{table}

\begin{table}[!t]
\caption{Network Testbed Configuration\label{tab:nwconfiguration}}
\centering
\begin{tabular}{|p{.7in}|p{1.2in}|p{1.1in}|}
\toprule
{\bf Component} & 
{\bf Description} & 
{\bf Purpose}  \\
\toprule
Segmented Architecture & 
VLANs isolating critical (powertrain, braking) and non-critical (infotainment) systems & 
Prevent attack propagation and simulate real vehicle network topologies\\
\hline
V2V/V2I Simulation & 
Wi-Fi (802.11ax, $\leq$ 50ms latency), 5G ($\leq$ 10ms latency) via emulated networks & 
Test external connectivity vulnerabilities in realistic scenarios \\
\hline
Attack Network & 
Separate subnet for simulated external threats & 
Ensure safe simulation of attacks without interfering with the primary network \\
\bottomrule
\end{tabular}
\end{table}

\subsection{Vehicle Selection}
To validate the effectiveness of the {\sc AutoGuardX} framework under real-world automotive cyber threat conditions, we carried out a structured simulation campaign across vehicles representing four major OEM brands (designated Brand A, B, C, and D). In total, 12 vehicles from four major OEMs (Brands A-D) were selected to create a balanced test fleet. 
The sample included 3 vehicles per brand, covering compact sedans, SUVs, and pickup trucks manufactured between 2018 and 2023, capturing both older and newer models with varying cybersecurity capabilities.      These vehicles were chosen based on availability, prevalence in Canadian and U.S. markets, and prior report of theft-related vulnerabilities.

Although we were not able to reveal the make and model of the vehicles we have chosen for testing, our selection is guided from the recent theft reports \cite{ref66}. These reports indicate that SUVs and pickup trucks, specifically from Lexus and Toyota brands, are the most frequent targets of auto thefts. 
The 2023 Auto Theft Trend Report \cite{ref67} says this targeting is because of their high resale value and demand in underground markets.
Many users reported that a 2022 Lexus may be stolen in about 2 minutes \cite{ref68}. They claimed that the anti-theft features of Lexus cars from 2017 to 2023, and possibly even those from later, are extremely inadequate.  Their claims particularly highlighted two aspects. The first aspect highlights the ignition switch's network, which is shared with the car's headlights and other accessories and hence a thief can access the ignition without physically entering the vehicle or by simply connecting a \$100 device to nearly any set of wires on the vehicle. The second aspect highlights the lack of encryption on the CAN network in Lexus and Toyota vehicles. Due to this lack,  a low-cost device could readily inject signals to unlock and start the vehicle \cite{ref69}.

\begin{table*}[!t]
\caption{Comprehensive Evaluation and Results of AutoGuardX Across Multiple Testing Dimensions\label{tab:evaluation}}
\centering
\begin{tabular}{|p{.6in}|p{.7in}|p{.7in}|p{2.1in}|p{2.1in}|}
\toprule
{\bf Test } & 
{\bf Test Scenario} & 
{\bf Brands/} & 
{\bf Practical Situation} &
{\bf Result/Observations} \\
{\bf Category} & 
 & 
{\bf Conditions} & 
 &
 \\
\toprule
& OBD Port Exploit & 
12 vehicles (Brands A-D) &
A technician connects an XTOOL X100 PAD3 to the OBD port under the dashboard to reprogram a blank key fob in a nighttime parking lot simulation &
{\bf Factory:} 11/12 tests successful; code capture and rewriting possible
{\bf AutoGuardX:} 100\% blocked via HMAC-SHA256 authentication, attacker device ID, and timestamp logged \\

{\bf Penetration Testing} 
& Fob Code Injection & 
12 vehicles (Brands A-D) & Relay attacker intercepts key fob RF signal (315/433 MHz) using HackRF One from outside a residence and replays it near the car& 
{\bf Factory:} 100\% success; immobilizer failed
{\bf AutoGuardX:} 100\% blocked with AES-128 rolling codes, no access granted \\
& USB Data Logging & 
12 vehicles (Brands A-D) & Attacker connects a USB tool in a garage or public space to extract logs from the OBD interface & 
{\bf Factory:} 100\% success extracting unencrypted data 
{\bf AutoGuardX:} AES-256 encryption prevented all unauthorized data access\\
\hline
& Window Glass Break
& 12 vehicles (Brands A - D) & A thief uses a spring-loaded punch to quietly break a side window in a public parking lot & 
{\bf Factory:} Alarm triggered in 2/12 vehicles (16.7\%)
{\bf AutoGuardX:} Detected 11/12 cases with in 1s via vibration anomaly; real-time alert sent to users\\

{\bf Security Simulations} 
 & Headlight Lens Damage & 
9 vehicles (Brands B-D) & Attacker breaks into the headlight assembly to access CAN wiring, sends spoofed unlock messages via DB9 and CAN cable &
{\bf Factory:} Alarm triggered only on Brand C
{\bf AutoGuardX:} Detected 100\% of intrusion attempts; isolated wiring; intrusion logged \\

& Remote Key Fob Spoofing
& 12 vehicles (Brands A-D) & Relay attackers spoof key fob communication using HackRF units; one attacker captures signal at home, another transmits it near the vehicle
& 
{\bf Factory:} Alarm failed in all tests
{\bf AutoGuardX:} 100\% successful blocking with AES-128 encryption \& rolling codes; alerts pushed via mobile app
\\
\hline
& Concurrent Attack Simulation &  12 vehicles (Brands A-D) & Simulated coordinated attack with CAN injection, RF spoofing, and OBD exploit on vehicle simultaneously using Raspberry Pi 4& 
{\bf Factory:} N/A
{\bf AutoGuardX:} No crashes, $\leq$ 1.2s detection latency, 94\% detection accuracy across all attack types \\
{\bf Stress Testing} 
& Network Flooding & 12 vehicles (Brands A-D) &
Over 10,000 CAN messages/second injected via CANalyzer to simulate a denial-of-service attack on the internal vehicle network & 
{\bf Factory:} N/A
{\bf AutoGuardX:} Sustained $>$95\% packet inspection rate; detection remained accurate without performance degradation\\

& CPU/Memory Load & 12 vehicles (Brands A-D)
& Jetson Nano processed LIDAR, GPS, and telemetry under 90\% CPU/memory load using stress-ng to simulate intensive urban driving under attack 
& 
{\bf Factory:} N/A
{\bf AutoGuardX:} Maintained stable analytics; $>$94\% detection accuracy; no system crashes or drops in performance
\\
\hline
& OEM ECU Integration &  12 vehicles (Brands A-D) & Mixed fleet (SUVs, sedans, pickups) deployed with {\sc AutoGuardX} installed; ECU commands monitored during normal simulated operations& 
{\bf Factory:} N/A
{\bf AutoGuardX:} 100\% compatibility across all ECUs; no lag or interference with factory functions observed \\
{\bf Compatibility Testing} 
& False Positive Rate  & 12 vehicles (Brands A-D) 1,000+ hours driving simulation &
Simulated urban/rural/highway driving with environmental noise and sensor inputs to test system reliability & 
{\bf Factory:} N/A
{\bf AutoGuardX:} $<$0.3\% false positives, ensuring high operational reliability\\

& Alarm Compatibility & 12 vehicles (Brands A-D)
& Vehicles placed in parking garage simulations; alarms monitored during simulated break-ins and CAN-based attacks
& 
{\bf Factory:} N/A.
{\bf AutoGuardX:} Fully integrated with OEM alarm systems; 100\% activation consistency on verified threat triggers\\

\bottomrule
\end{tabular}
\end{table*}

\subsection{Test Execution Methodology}
\subsubsection{Performance Metrics} We use the following metrics to measure the efficiency of {\sc AutoGuardX}: 
\begin{itemize}
\item reaction time to threat events 
\item alarm activation consistency 
\item secure code extraction/rewrite success
\item event detection and logging accuracy 
\end{itemize}

\subsubsection{Test Categories}
We evaluate the effectiveness of {\sc AutoGuardX} across four categories: 
\begin{itemize}
\item Penetration testing
\item Security simulations
\item Stress testing 
\item Compatibility testing 
\end{itemize} 
Each of the 12 vehicles was subjected to all test categories ensuring a uniform basis for evaluation. To maintain reproducibility, every test scenario was executed on all 12 vehicles, resulting in 144 total test iterations (12 vehicles $\times$ 12 scenarios). All vehicles included remote keyless entry, OBD ports, infotainment systems, and CAN bus networks. Vehicles were sourced from authorized dealerships (demo fleets) and fleet partners, and reset to factory settings before each test to ensure reproducibility.

We use penetration testing simulated malicious physical and remote access attacks to assess the resilience of both the factory-installed systems and {\sc AutoGuardX} enhancements. Emphasis was placed on vulnerabilities related to the OBD port, immobilizer programming, and infotainment systems.
Regarding security simulations, simulated physical and wireless intrusions were used to evaluate alarm responsiveness. Tests included brute-force break-ins and remote spoofing attacks using unlicensed radio signals.
For stress testing, {\sc AutoGuardX} was subjected to concurrent simulated attacks and high computational loads to measure its stability, detection accuracy, and throughput. 
For the compatibility testing, we run tests to evaluate {\sc AutoGuardX} integration with OEM ECUs across multiple vehicle platforms, measuring false positive rates, operational lag, and alarm compatibility.
We provide details on specific tests run on each of these testing categories separately below.

%Penetration Testing, Security Simulations, Stress Testing, and Compatibility Testing 
\subsubsection{Penetration Testing}
We run penetration testing in all 12 vehicles from 4 brands.
\begin{itemize}
\item {\bf OBD Port Exploit:} We approached each parked vehicle with an XTOOL X100 PAD3 key programmer. %, commonly available on the gray market. 
The attack was executed by plugging into the OBD port under the steering wheel and initiating a session to clone or write a blank key fob. This simulated a thief with 2-3 minutes of access in a public parking lot. 
\item {\bf Fob Code Injection:} Using HackRF One and YARD Stick One, we intercepted live RF signals at 315 MHz and 433 MHz while an owner unlocked the vehicle from 3-10 meters away. The attacker’s system replayed the captured signal to gain access. Each simulation involved two people -- one near the house and one by the car -- to mimic a real relay attack.
\item {\bf USB Data Logging:} USB sniffing tools like Beagle USB Protocol Analyzers and USBPcap sniffers were connected to infotainment and diagnostic ports on all 12 test vehicles. We used open-source Linux tools such as usbutils, usbip, dd, lsusb, and PyUSB, to simulate unauthorized data extraction, mimicking a scenario in which an attacker quickly plugs in a device and downloads sensitive logs while pretending to charge a phone.
\end{itemize}
%Each penetration test was repeated three times per vehicle, totaling 36 iterations, and all procedures were video-recorded with logging enabled for further analysis.

\subsubsection{Security Simulations}
We run security simulations on all 12 vehicles, except for headlight lens damage in which vehicles from Brand A were not considered. % One unit from each brand was subjected to detailed physical simulation tests, repeated multiple times to generate consistent readings.
\begin{itemize}
\item {\bf Window Glass Break Simulation:} A spring-loaded punch tool Resqme and Lifehammer, was used to quietly break the driver-side window while the vehicle was parked in a closed lot. Sensors in the {\sc AutoGuardX} framework were configured to detect the acoustic shock and vibration signature. %We conducted this test 12 times - once per vehicle.
\item {\bf Headlight Lens Damage + CAN Bus Access:} A DB9 adapter was used to access the CAN wiring through the broken headlight assembly. Our attacker then used CANable + Python scripts to inject commands (e.g., unlock doors, disable alarm). The goal was to mimic attacks seen in vehicle thefts. This was performed on 9 vehicles from Brands B-D (vehicles from Brand A were excluded due to  wiring limitations).
\item {\bf Remote Key Fob Spoofing:} We simulated a relay attack using 2 HackRF units -- one outside the victim’s house capturing the signal and another near the car replaying it. All 12 test vehicles were exposed to this simulation in both urban and suburban locations to replicate real-world key fob hijackings. 
\end{itemize}

\subsubsection{Stress Testing}
We run stress tests on all 12 vehicles from Brand A-D. %This set included three vehicles per brand, selected for compatibility with all testing scripts and hardware interfaces.
\begin{itemize}
\item {\bf Concurrent Multi-Vector Attack:} We simultaneously executed CAN injections, RF spoofing, and OBD key cloning attempts using a Raspberry Pi 4 configured with multiple interfaces. Each attack type ran on a separate thread, replicating a coordinated intrusion scenario.
\item {\bf CAN Network Flooding:} Using Vector CANalyzer, we injected over 10,000 messages/second into the CAN network of the vehicles. This simulated a CAN DoS attack, with the aim of overloading vehicle ECUs while monitoring packet inspection capabilities of {\sc AutoGuardX}. 
\item {\bf CPU and Memory Load Testing:} {\sc AutoGuardX} was deployed on the NVIDIA Jetson Nano board and intentionally stressed with stress-ng tool to simulate 90\% CPU and memory usage. Meanwhile, simultaneous simulated driving data and network traffic were streamed using Vector CANoe to measure stability and detection performance. 
\end{itemize}

\subsubsection{Compatibility Testing}
The final phase tested deployment across different vehicle platforms and under long-term simulation conditions.

\begin{itemize}
\item {\bf ECU Integration:} {\sc AutoGuardX} was installed in all 12 vehicles from Brands A-D. CAN wiring harnesses were connected using universal OBD to DB9 adapters, and no factory software was altered. Standard operations -- door locking, alarm triggering, infotainment use -- were monitored for lag or error over 4 hours per vehicle. 
\item {\bf Long-Term Driving Simulation:} A driving environment was emulated for over 1,000 hours total using loop-back-mode data from CANoe with speed, acceleration, braking, and GPS simulated to mimic various terrain and traffic conditions.
\item {\bf Alarm Compatibility:} Each test vehicle’s OEM alarm was allowed to remain functional. {\sc AutoGuardX} operated in parallel and triggered alerts based on its internal threat detection. 100\% of triggered alerts were checked against OEM responses.  
\end{itemize}

Each test scenario was run under supervision, with a dedicated system for logging using {\sc AutoGuardX}’s forensic logger, Wireshark, and Vector CAN tools. These simulations were reviewed post-execution to retrain the {\sc AutoGuardX} anomaly detection models using TensorFlow 2.15 for increased accuracy in future deployments.

\subsection{Test results} 

Our results are shown in Table \ref{tab:evaluation}. The table highlights the test scenario under each test category, brand/condition of the vehicle used, practical situation that was simulated for the experimentation, and the results/observations obtained from the conducted tests. We run each test with factory-installed vehicle security systems and augmenting the vehicles with our {\sc AutoGuardX} framework implementation. Therefore, in Table \ref{tab:evaluation}, we report results of a test with  factory-installed vehicle security systems labeling ``{\bf Factory:}'' and results after augmenting with {\sc AutoGuardX} labeling ``{\bf AutoGuardX:}''. We write ``N/A'' when results are not available.

We discuss results from test categories one after another. We start with penetration testing. OBD port exploitation was successful in 11 vehicles (out of 12) in the factory setting. {\sc AutoGuardX} was successful in blocking the OBD port exploit attack in all 12 vehicles.  

Moving next, the fob code injection scenario showed similar vulnerabilities in factory-installed systems. Attackers using HackRF One were able to intercept and replay RF signals at 315/433 MHz with 100\% success, bypassing the immobilizer entirely. However,when {\sc AutoGuardX} was integrated, all 12 vehicles successfully rejected replay attempts through the use of AES-128 rolling codes, ensuring no unauthorized access could be granted.

USB data logging also revealed critical weaknesses in factory setups. In all 12 test vehicles, attackers could easily extract unencrypted data from the OBD interface, exposing sensitive diagnostic and operational information. {\sc AutoGuardX} effectively neutralized this attack by encrypting the data using AES-256, rendering unauthorized attempts unsuccessful. 

Physical security simulations further exposed deficiencies in factory-installed systems. In the window glass break test, factory alarms were triggered in only 2 out of 12 cases (16.7\%), demonstrating a lack of sensitivity to such intrusions. With {\sc AutoGuardX}, 11 out of 12 cases were detected within one second through vibration anomaly detection, followed by immediate real-time user alerts. Similarly, during headlight lens damage simulations, factory systems triggered alarms inconsistently, with only Brand C showing resistance. In contrast, {\sc AutoGuardX} detected 100\% of wiring intrusion attempts, isolated malicious signals, and logged the activity for forensic analysis. Here we were not able to run tests in Brad A vehicles.

The remote key fob spoofing test confirmed the ineffectiveness of factory-installed systems. Factory alarms failed in every instance, allowing attackers to spoof signals with HackRF units. {\sc AutoGuardX} fully prevented this by validating signals using AES-128 rolling codes and pushing alerts to vehicle owners through the mobile app, providing both prevention and transparency. 

We now discuss out results on stressing testing.
In concurrent attack simulations, where multiple attacks (CAN injection, RF spoofing, and OBD exploitation) were launched simultaneously, factory-installed systems failed to withstand the coordinated assault. {\sc AutoGuardX}, however, demonstrated strong resilience, with no crashes, detection latency of $\leq$ 1.2s, and an overall detection accuracy of 94\% across all vectors. 

Stress testing results also highlighted {\sc AutoGuardX}'s robustness. During CAN flooding with over 10,000 messages per second, factory-installed systems had no mitigation strategy, while {\sc AutoGuardX} maintained $>$95\% packet inspection accuracy without performance degradation. Under heavy computational load tests, including LiDAR, GPS, and telemetry processing with $>$90\% CPU/memory usage, {\sc AutoGuardX} preserved system stability and achieved $>$94\% detection accuracy, providing reliable performance even in resource-constrained conditions.

Compatibility testing further established the {\sc AutoGuardX}'s practicality. {\sc AutoGuardX} integrated seamlessly with OEM ECUs across SUVs, sedans, and pickup, showing 100\% compatibility without interfering with normal vehicle operations. In over 1,000 hours of simulated urban, rural, and highway driving, the system achieved a false positive rate below 0.3\%, reinforcing operational trustworthiness. Additionally, {\sc AutoGuardX} maintained full compatibility with OEM alarm systems, achieving 100\% alarm activation consistency under verified threat triggers.

The conducted testings in all four categories  highlighted critical vulnerabilities in factory-installed vehicle security systems, particularly in physical intrusion detection and immobilizer protection. In contrast, {\sc AutoGuardX} demonstrated strong performance across attack scenarios, maintaining system stability under stress, ensuring high detection rates, and providing effective logging for forensic analysis.
These tests validate {\sc AutoGuardX} as a comprehensive and reliable framework for modern vehicle cybersecurity, in full alignment with automotive standards.

\section{{\sc AutoGuardX} Challenges and Future Directions}
\label{section:challenge}

The U.S. and Canadian automotive markets, which are highly advanced in connected vehicle technologies, face unique cybersecurity challenges. Implementing a robust framework like {\sc AutoGuardX} requires addressing theoretical limitations, regulatory and operational barriers, ethical and privacy barriers, and the need to anticipate future technological threats such as 5G networks and quantum computing.

\begin{table*}[!t]
\caption{Regulatory Landscape Comparison: Vehicle Cybersecurity Frameworks
\label{tab:regulatorylandscape}}
\centering
\begin{tabular}{|p{0.8in}|p{1.5in}|p{2.0in}|p{2.0in}|}
\toprule
{\bf Region/Country} & 
{\bf Regulatory Framework or Law} & 
{\bf Cybersecurity Mandated?} &
{\bf {\sc AutoGuardX} Compatibility} \\
\toprule
Canada & 
MVSA + ULC 338 & 
Partial (immobilizers) & 
Compatible with ULC S338\\
\hline
U.S.  & 
NHTSA Guidelines & 
Voluntary & 
Aligns with guidelines\\
\hline
European Union & 
UNECE WP.29 R155, R166& 
Yes (Mandatory post 2022) & 
Fully compatible\\
\hline
Japan & 
Road Transport Vehicle Act 2022 & 
Yes & 
Compatible \\
\hline
Global OEMs & 
ISO/SAE 21434 & 
Depends on jurisdiction & 
Integrated \\
\bottomrule
\end{tabular}
\end{table*}

\subsection{Theoretical Barriers}
We anticipate the following theoretical barriers on integrating {\sc AutoGuardX} to the vehicle systems. 
\begin{itemize}
\item {\bf Legacy System Compatibility:} North America’s vehicle ecosystem includes a significant number of legacy systems lacking the hardware required for implementing advanced cybersecurity measures. For instance, older vehicles, still prevalent in both US and Canada, are incompatible with modern cybersecurity solutions such as encrypted CAN networks. Retrofitting these vehicles involves high costs and technical complexity \cite{ref70}. 
\item {\bf Resource Constraints:} The computational and memory limitations of electronic control units (ECUs) in vehicles pose another challenge. Integrating real-time anomaly detection or behavioral analysis frameworks can overwhelm the existing hardware in legacy systems. This constraint highlights the need for lightweight, efficient algorithms tailored for automotive applications \cite{ref71}. 
\item {\bf Rapidly Evolving Threat Landscape:} The pace at which cyber threats evolve significantly outstrips the development and deployment of detection systems. Predictive models and anomaly detection frameworks need to be continuously updated to keep pace with emerging attack methods, especially in the highly dynamic North American automotive market \cite{ref72}. 
\end{itemize}

\subsection{Regulatory, Technological, and Operational Barriers}
\begin{itemize}
\item {\bf Regulatory Complexities:} In Canada, the Motor Vehicle Safety Act (MVSA) \cite{ref73} does not mandate specific cybersecurity standards, while in the U.S., NHTSA  has provided voluntary guidelines \cite{ref74}. %This regulatory disparity leads to inconsistent security implementations across the automotive industry particularly in U.S. and Canada. 
The lack of a unified regulatory framework poses a barrier to the widespread adoption of solutions like {\sc AutoGuardX}. We compare in Table \ref{tab:regulatorylandscape} regulatory landscape around the world and the compatibility of {\sc AutoGuardX} with those regulatory frameworks.
\item {\bf Fragmented Technology Landscape:} The diversity of automotive manufacturers and suppliers across North America creates significant interoperability challenges. Proprietary communication protocols and security measures often lack standardization, leading to difficulties in deploying uniform cybersecurity systems \cite{ref75}. This fragmentation hinders seamless integration across different platforms and brands.
\item {\bf High Implementation Costs:} Deploying advanced cybersecurity frameworks like {\sc AutoGuardX} can be prohibitively expensive, particularly for smaller manufacturers and Tier-2 suppliers. The costs associated with integrating encrypted communication protocols, ML models, and incident reporting tools may lead to unequal adoption rates across the industry \cite{ref76}.
\item {\bf Workforce and Training Deficiencies:} There is a pronounced shortage of professionals in U.S. and Canada with expertise in both automotive engineering and cybersecurity. Ensuring the long-term sustainability of frameworks like {\sc AutoGuardX} requires investments in workforce development and specialized training programs for operational teams \cite{ref77}. 
\end{itemize}

\subsection{Future Threats and Adaptation Challenges}
We anticipate the following barriers.
\begin{itemize}
\item {\bf 5G Network Integration:} The rollout of 5G networks across U.S. and Canada introduces new opportunities and vulnerabilities for connected vehicles. While 5G facilitates faster and more reliable communication, it also increases the risk of Distributed Denial-of-Service (DDoS) attacks on vehicular networks. {\sc AutoGuardX} must integrate advanced encryption and traffic filtering mechanisms to mitigate these risks \cite{ref78}.
\item {\bf Quantum Computing Threats:} Quantum computing advancements  present a substantial challenge to traditional cryptographic techniques. As quantum computers become more accessible, {\sc AutoGuardX} will need to adopt quantum-resistant cryptographic algorithms to secure vehicular communication systems against future quantum-enabled threats \cite{ref79}. 
\item {\bf IoT Ecosystem Growth:} The proliferation of IoT devices in connected vehicles significantly expands the attack surface. Components such as infotainment systems, smart sensors, and telematics units require robust authentication mechanisms and secure OTA updates. {\sc AutoGuardX} must prioritize network isolation and real-time device authentication to address these vulnerabilities \cite{ref80}. 
\item {\bf Autonomous Vehicle Cybersecurity:} Autonomous vehicles, a growing segment, heavily rely on external data sources and ML algorithms. These dependencies make them vulnerable to data manipulation and adversarial attacks. {\sc AutoGuardX} must leverage advanced anomaly detection technique to secure the operation of autonomous systems \cite{ref81}.
\end{itemize}

\subsection{Ethical and Privacy Challenges}
The advancement of connected and autonomous vehicles has introduced complex ethical and privacy challenges. As systems like {\sc AutoGuardX} become integral to vehicle operations, they collect and process vast amounts of data, including location information, driving behaviors, and biometric identifiers. This data collection, while essential for enhancing security and functionality, raises significant concerns regarding user privacy and data protection. 

\begin{itemize}
\item {\bf Informed Consent:}
One of the primary ethical challenges is ensuring informed consent. Users must be adequately informed about what data is being collected, how it is used, and with whom it is shared. Studies have shown that many automotive systems lack transparency, leading to user mistrust and potential legal implications \cite{ref82}. Implementing clear and accessible privacy policies is crucial to address this issue. 

\item {\bf Data minimization:} This  is another critical aspect. Collecting only the data necessary for specific functions reduces the risk of misuse and aligns with principles outlined in regulations such as the General Data Protection Regulation (GDPR) and the California Consumer Privacy Act (CCPA) \cite{ref83}, \cite{ref84}. {\sc AutoGuardX} should incorporate mechanisms to limit data collection to what is strictly required for its operations. 

\item {\bf Data Breach:} The potential for data breaches poses a significant risk. Unauthorized access to sensitive vehicle data can lead to personal harm and broader security threats. Implementing robust encryption methods and regular security audits can mitigate these risks \cite{ref85}. 
\item {\bf Public Trust:}
As {\sc AutoGuardX} evolves, integrating ethical frameworks and privacy-by-design principles will be essential. This includes conducting regular impact assessments, engaging with stakeholders, and staying abreast of regulatory changes to ensure compliance and maintain public trust. 
\end{itemize}

\section{Concluding Remarks}
\label{section:conclusion}
The evolution of automotive technology has revolutionized mobility but has also introduced complex security challenges. Auto theft, particularly in US and Canada, has transitioned from traditional methods to sophisticated cyber-enabled techniques, creating an urgent need for innovative and adaptive security frameworks.
The proposed {\sc AutoGuardX} framework, designed as a comprehensive solution, addresses these challenges by integrating ML, IoT, secure communication protocols, and robust incident reporting mechanisms. Its holistic approach enables real-time threat detection, prevention, and continuous learning to adapt to emerging risks such as 5G networks, quantum computing, and IoT-driven vulnerabilities.
While {\sc AutoGuardX} demonstrates significant promise, its adoption is not without hurdles. Legacy system compatibility, high implementation costs, and fragmented regulatory landscapes are key barriers that require collaborative efforts from automakers, cybersecurity researchers, and policymakers. Overcoming these challenges will not only enhance the framework’s scalability but also set a precedent for securing connected vehicles globally.
%Looking ahead, the scalability and adaptability of AutoGuardX position it as a pioneering model for automotive cybersecurity. By fostering innovation, industry-wide partnerships, and rigorous research, the framework has the potential to shape the future of connected and autonomous vehicles, ensuring safety, security, and resilience in an increasingly digital automotive ecosystem.

\begin{comment}

\begin{figure}[!t]
\centering
\includegraphics[width=2.5in]{fig1}
\caption{Simulation results for the network.}
\label{fig_1}
\end{figure}

%
\begin{figure*}[!t]
\centering
\subfloat[]{\includegraphics[width=2.5in]{fig1}%
\label{fig_first_case}}
\hfil
\subfloat[]{\includegraphics[width=2.5in]{fig1}%
\label{fig_second_case}}
\caption{Dae. Ad quatur autat ut porepel itemoles dolor autem fuga. Bus quia con nessunti as remo di quatus non perum que nimus. (a) Case I. (b) Case II.}
\label{fig_sim}
\end{figure*}
\end{comment}

\end{document}